\documentstyle[12pt,epsf]{article}

\newlength{\pubnumber} 

\catcode`\@=11
\@addtoreset{equation}{section}

\def\section{\@startsection{section}{1}{\z@}{3.5ex plus 1ex minus .2ex}
 {2.3ex plus .2ex}{\large\bf}}
\def\subsection{\@startsection{subsection}{2}{\z@}{2.3ex plus .2ex}
 {2.3ex plus .2ex}{\bf}}

\begin{document}

\begin{titlepage}
\samepage{
\setcounter{page}{1}
\rightline{ACT-11/98}
\rightline{CTP-TAMU-45/98}
\rightline{TPI-MINN-98/24}
\rightline{UMN-TH-1729-98}
\rightline{\tt hep-ph/9811427}
\rightline{November 1998}
\vfill
\begin{center}
 {\Large \bf  String Derived MSSM and M--theory Unification\\ }
\vspace{.35truecm}
\vfill {\large
        G.B. Cleaver,$^{1,2}$\footnote{gcleaver@rainbow.physics.tamu.edu}
        A.E. Faraggi,$^{3}$\footnote{faraggi@mnhepw.hep.umn.edu}
        and
        D.V. Nanopoulos$^{1,2,4}$\footnote{dimitri@soda.physics.tamu.edu}}
\\
\vspace{.1truecm}
{\it $^{1}$ Center for Theoretical Physics,
            Dept.\  of Physics, Texas A\&M University,\\
            College Station, TX 77843, USA\\}
\vspace{.04in}
{\it $^{2}$ Astro Particle Physics Group,
            Houston Advanced Research Center (HARC),\\
            The Mitchell Campus,
            Woodlands, TX 77381, USA\\}
\vspace{.04in}
{\it$^{3}$  Department of Physics, University of Minnesota, 
            Minneapolis, MN 55455, USA\\}
\vspace{.02in}
{\it$^{4}$  Academy of Athens, Chair of Theoretical Physics, 
            Division of Natural Sciences,\\
            28 Panepistimiou Avenue, Athens 10679, Greece\\}
%\vspace{.025in}
\end{center}
%\vfill
\begin{abstract}
The recent conjecture of possible equivalence between the
string scale $M_S$ and the minimal supersymmetric standard model 
unification scale $M_U\approx 2.5\times 10^{16}$ GeV 
is considered in the context of string models. This 
conjecture suggests that the observable gauge group 
just below the string scale should be 
$SU(3)_C\times SU(2)_L\times U(1)_Y$ and that the 
$SU(3)_C\times SU(2)_L\times U(1)_Y$--charged spectrum of the 
observable sector should consist solely of the MSSM spectrum.
We demonstrate that string models can actually be constructed
that possess these observable features. 
Two aspects generic to many classes of 
three family $SU(3)_C\times SU(2)_L\times U(1)_Y$ string 
models are both an extra local anomalous $U(1)_A$ and 
numerous (often fractionally charged) exotic particles beyond
the MSSM.  
Thus, for these classes, the key to obtaining an
$M_S = M_U\approx 2.5\times 10^{16}$ GeV 
string model is the existence of $F$-- and $D$--flat directions that  
near the string scale can
simultaneously break the anomalous $U(1)$ and 
give mass to {\it all} exotic SM--charged 
observable particles,  
decoupling them from the low energy spectrum.
In this letter we show, in the context of free fermionic strings,
that string models with flat directions possessing these 
features do exist. We present one such string derived model
in which all such exotic observable states beyond the MSSM receive
mass at the scale generated by the Fayet--Iliopoulos term.
The associated $F$-- and $D$--flat direction is proven flat 
to all orders of the superpotential.
\end{abstract}
\begin{center}
{\it To appear in Physics Letters B.}
\end{center}}
\end{titlepage}

\setcounter{footnote}{0}

% ========================= DEFINITIONS ===================================

\def\beq{\begin{equation}}
\def\eeq{\end{equation}}
\def\beqn{\begin{eqnarray}}
\def\eeqn{\end{eqnarray}}

\def\dag{\dagger}
\def\qandq{\quad {\rm and} \quad} 
\def\qand{\quad {\rm and} } 
\def\andq{ {\rm and} \quad } 
\def\qwithq{\quad {\rm with} \quad} 
\def\qwith{ \quad {\rm with} } 
\def\withq{ {\rm with} \quad} 

\def\no{\noindent }
\def\nolabel{\nonumber }
\def\ie{i.e., }
\def\eg{{\it e.g.}}
\def\half{{\textstyle{1\over 2}}}
\def\third{{\textstyle {1\over3}}}
\def\quarter{{\textstyle {1\over4}}}
\def\sixth{{\textstyle {1\over6}}}
\def\m{{\tt -}}
\def\p{{\tt +}}

\def\Tr{{\rm Tr}\, }
\def\tr{{\rm tr}\, }

\def\MP{M_{P}}
\def\hb{\bar{h}}
\def\slash#1{#1\hskip-6pt/\hskip6pt}
\def\slk{\slash{k}}
\def\GeV{\,{\rm GeV}}
\def\TeV{\,{\rm TeV}}
\def\y{\,{\rm y}}
\def\SM{Standard--Model }
\def\SUSY{supersymmetry }
\def\SSSM{supersymmetric standard model}
\def\MSSM{minimal supersymmetric standard model}
\def\smgg{ $SU(3)_C\times SU(2)_L\times U(1)_Y$ }
\def\vev#1{\left\langle #1\right\rangle}

\def\l{\langle}
\def\r{\rangle}
\def\o#1{\frac{1}{#1}}

\def\Htw{{\tilde H}}
\def\chibar{{\overline{\chi}}}
\def\qbar{{\overline{q}}}
\def\ibar{{\overline{\imath}}}
\def\jbar{{\overline{\jmath}}}
\def\Hbar{{\overline{H}}}
\def\Qbar{{\overline{Q}}}
\def\abar{{\overline{a}}}
\def\alphabar{{\overline{\alpha}}}
\def\betabar{{\overline{\beta}}}
\def\tautwo{{ \tau_2 }}
\def\thetatwo{{ \vartheta_2 }}
\def\thetathree{{ \vartheta_3 }}
\def\thetafour{{ \vartheta_4 }}
\def\ttwo{{\vartheta_2}}
\def\tthree{{\vartheta_3}}
\def\tfour{{\vartheta_4}}
\def\ti{{\vartheta_i}}
\def\tj{{\vartheta_j}}
\def\tk{{\vartheta_k}}
\def\calF{{\cal F}}
\def\smallmatrix#1#2#3#4{{ {{#1}~{#2}\choose{#3}~{#4}} }}
\def\ab{{\alpha\beta}}
\def\Minv{{ (M^{-1}_\ab)_{ij} }}
\def\ii{{(i)}}
\def\V{{\bf V}}
\def\N{{\bf N}}

% for basis vectors:
\def\b{{\bf b}}
\def\S{{\bf S}}
\def\X{{\bf X}}
\def\I{{\bf I}}
\def\bone{{\mathbf 1}}
\def\bs{{\mathbf S}}
\def\bb{{\mathbf b}}
\def\mb{{\mathbf b}}
\def\mS{{\mathbf S}}
\def\bS{{\mathbf S}}
\def\bs{{\mathbf S}}
\def\mX{{\mathbf X}}
\def\mI{{\mathbf I}}
\def\bI{{\mathbf I}}
\def\balpha{{\mathbf \alpha}}
\def\bbeta{{\mathbf \beta}}
\def\bgamma{{\mathbf \gamma}}
\def\bxi{{\mathbf \xi}}
\def\malpha{{\mathbf \alpha}}
\def\mbeta{{\mathbf \beta}}
\def\mgamma{{\mathbf \gamma}}
\def\mxi{{\mathbf \xi}}

\def\eps{\epsilon}

\def\t#1#2{{ \Theta\left\lbrack \matrix{ {#1}\cr {#2}\cr }\right\rbrack }}
\def\C#1#2{{ C\left\lbrack \matrix{ {#1}\cr {#2}\cr }\right\rbrack }}
\def\tp#1#2{{ \Theta'\left\lbrack \matrix{ {#1}\cr {#2}\cr }\right\rbrack }}
\def\tpp#1#2{{ \Theta''\left\lbrack \matrix{ {#1}\cr {#2}\cr }\right\rbrack }}
\def\l{\langle}
\def\r{\rangle}

%================== BLACKBOARD BOLD CHARACTERS ==============================

\def\inbar{\,\vrule height1.5ex width.4pt depth0pt}

\def\IC{\relax\hbox{$\inbar\kern-.3em{\rm C}$}}
\def\IQ{\relax\hbox{$\inbar\kern-.3em{\rm Q}$}}
\def\IR{\relax{\rm I\kern-.18em R}}
 \font\cmss=cmss10 \font\cmsss=cmss10 at 7pt
\def\IZ{\relax\ifmmode\mathchoice
 {\hbox{\cmss Z\kern-.4em Z}}{\hbox{\cmss Z\kern-.4em Z}}
 {\lower.9pt\hbox{\cmsss Z\kern-.4em Z}}
 {\lower1.2pt\hbox{\cmsss Z\kern-.4em Z}}\else{\cmss Z\kern-.4em Z}\fi}

%========================================================================
%          MACROS FOR REFERENCES
%========================================================================
\def\AEF{A.E. Faraggi}
\def\NPB#1#2#3{{\it Nucl.\ Phys.}\/ {\bf B#1} (19#2) #3}
\def\NPBPS#1#2#3{{\it Nucl.\ Phys.}\/ {{\bf B} (Proc. Suppl.) {\bf #1}} (19#2) 
 #3}
\def\PLB#1#2#3{{\it Phys.\ Lett.}\/ {\bf B#1} (19#2) #3}
\def\PRD#1#2#3{{\it Phys.\ Rev.}\/ {\bf D#1} (19#2) #3}
\def\PRL#1#2#3{{\it Phys.\ Rev.\ Lett.}\/ {\bf #1} (19#2) #3}
\def\PRT#1#2#3{{\it Phys.\ Rep.}\/ {\bf#1} (19#2) #3}
\def\MODA#1#2#3{{\it Mod.\ Phys.\ Lett.}\/ {\bf A#1} (19#2) #3}
\def\IJMP#1#2#3{{\it Int.\ J.\ Mod.\ Phys.}\/ {\bf A#1} (19#2) #3}
\def\nuvc#1#2#3{{\it Nuovo Cimento}\/ {\bf #1A} (#2) #3}
\def\RPP#1#2#3{{\it Rept.\ Prog.\ Phys.}\/ {\bf #1} (19#2) #3}
\def\etal{{\it et al\/}}

%==============================================================================
\hyphenation{su-per-sym-met-ric non-su-per-sym-met-ric}
\hyphenation{space-time-super-sym-met-ric}
\hyphenation{mod-u-lar mod-u-lar--in-var-i-ant}
%==============================================================================

%============================== SECTION 1 ============================

\setcounter{footnote}{0}
\section{Introduction}
One of the intriguing hints for new physics beyond the Standard
Model (SM) is the nearly perfect unification of the Standard Model
gauge couplings, assuming the spectrum of the Minimal Supersymmetric
Standard Model (MSSM) above the electroweak scale.
This coupling unification occurs at the scale of order
$M_{\rm MSSM}\approx 2\times10^{16}{\rm GeV}$, and is one or
two orders of magnitude below the scale where the gravitational
interaction becomes comparable in strength to the gauge interactions.
Gravity, on the other hand, can be unified consistently
with the gauge interactions only if the MSSM is 
embedded in superstring/M--theory. However, a general prediction
of perturbative string theory is that the gauge couplings
unify at a scale $M_S$ of the order of $5\times10^{17}{\rm GeV}$. 
Thus, at least in the context of perturbative string theory
it is natural to seek string models that contain the MSSM
spectrum plus additionally a few color and electroweak states
in vector--like representations. Such states can then receive
intermediate mass scale, and elevate the coupling unification
scale to the string scale
while keeping the MSSM running coupling strengths in agreement 
with their experimentally measured $M_{Z^o}$ scale values. 
 However, while string derived models
that allow such scenarios have indeed been constructed,
one may argue that the additional mass scales
require additional ad hoc fine tuning, and therefore
is not very attractive. 

It is then quite
remarkable that within the context of M--theory
a mechanism has been proposed that allows
the gauge couplings to unify at $\sim 10^{16}{\rm GeV}$, \cite{mth}
thus maintaining the successful MSSM prediction. Alas, 
from the point of view of M--theory unification
this would require that the observable gauge group
after compactification is precisely
$SU(3)_C\times SU(2)_L\times U(1)_Y$,
which means that there should be no GUT or semi--GUT
below the compactification (i.e., string) scale.
Furthermore, it implies that soon below this scale 
the (SM--charged) 
spectrum in the observable sector\footnote{By observable sector we
mean the part which affects the Standard Model gauge coupling
evolution, thus allowing the existence of unbroken
horizontal $U(1)$ symmetries, which do not mix with the Standard
Model gauge group. Such symmetries do not survive to low energies
in the model that we discuss, but it is a possibility
that may occur a priori.} must consist
solely of the MSSM spectrum, {\it i.e.}
three generations plus two electroweak Higgs doublets.
However, string models
generically give rise to several Higgs multiplets and
to many exotic states, charged under the Standard
Model, and which may remain massless at the string scale.
In fact, to date it is in general claimed that there does not
exist a single string model that produces solely
the MSSM spectrum in the observable sector below the string scale.

In this letter we therefore undertake the task
of deriving a string model that produces only the MSSM
spectrum in the $SU(3)_C\times SU(2)_L\times U(1)_Y$--charged 
observable sector.
The model that we study (referred to henceforth as the ``FNY model'')
is the string model of \cite{fny}
and was constructed in the free fermionic formulation.
Like all other string models which are constructed
by utilizing characters of level one Kac--Moody current
algebra, it contains exotic fractionally charged states
in the massless spectrum. As pointed out in ref.\ \cite{fc},
in this model all the exotic fractionally charged states
couple to a set of Standard Model singlets at the
cubic level of the superpotential. Thus, by assigning
vacuum expectation values (VEVs) 
to this set of Standard Model singlets
all the exotic fractionally charged states
receive mass of the order of the string scale
and decouple from the massless spectrum.

The model also contains, at the massless string level,
a number of electroweak Higgs doublets and 
a color triplet/anti--triplet pair beyond the MSSM. 
We show that by the same suitable choice
of flat directions that only one Higgs pair remains light below
the string scale. The additional color triplet/anti--triplet pair 
receives mass from a fifth order superpotential term. 
This results in the triplet pair receiving a mass that is 
slightly below the string scale and is perhaps smaller 
than the doublet and fractional exotic masses 
by a factor of around $(1/10-1/100)$.

Thus, all $SU(3)_C\times SU(2)_L\times U(1)_Y$--charged non--MSSM 
states decouple from the massless spectrum
at or slightly below the string scale by coupling to flat direction VEVs. 
Thus, for the first time we
present a string model that yields below the string
scale solely the spectrum of the MSSM in the observable 
sector.
Such a string model can therefore serve as an example
which satisfies the requirements imposed
by the M--theory motivated unification. 

Another interesting property of the FNY model
is that provided that the weak--hypercharge 
is left unbroken at the string scale,
the additional $U(1)_{Z^\prime}$ which
is embedded in $SO(10)$ is necessarily broken
by non--Abelian singlet flat directions near the string scale.
Thus, in this model the $SO(10)$ subgroup
below the string scale is necessarily $SU(3)\times SU(2)\times U(1)_Y$,
rather than $SU(3)\times SU(2)\times U(1)_C\times U(1)_L$,
which is another appealing property from the point of view
of M--theory motivated unification. 

There is a crucial new input \cite{cceel2}, which we now discuss,
that allows us to achieve, at present, our goal of 
a string--derived MSSM, 
nearly a decade after the original model was constructed. 
Let us recall that this string model contains an anomalous $U(1)$ 
symmetry \cite{anoma}  
(a generic feature of three family \smgg string models
of free fermionic, bosonic lattice, or orbifold construction).
Elimination of the anomalous $U(1)$ symmetry generates a 
Fayet--Iliopoulos (FI) term, 
by the VEV of the dilaton field, that breaks supersymmetry near the string 
scale.
To preserve supersymmetry near the string scale, one must satisfy $F$-- and 
$D$--flatness constraints arising from the superpotential, by giving VEVs to 
a set of standard model scalar singlets in the massless spectrum of the string 
models \cite{u1a}. Since these fields are typically also
charged under the non--anomalous gauge symmetries, a non--trivial set
of constraints is imposed on the possible choices of VEVs
and in general will break some or all of these symmetries
spontaneously. It turns out that in the FNY model of ref.\  \cite{fny}
the D--flatness constraints are particularly restrictive in the following
sense. In many free fermionic models utilizing the
NAHE set of boundary condition basis vectors,
one can solve the $D$--term constraints by assigning VEVs
solely to $SO(10)$ singlet fields \cite{so10sig,eu}. 
However, it turns out,
as we discuss below, that this is not possible in the model
of ref.\  \cite{fny}. In the FNY model one must assign
non--trivial VEVs to states which are Standard Model
singlets, but that are charged under the $U(1)_{Z^\prime}$
which is embedded in $SO(10)$, thus complicating the problem
of finding FNY flat directions. In fact,
the original motivation to construct the model of ref.\  \cite{eu}
was precisely to find flat $F$ and $D$ directions in the free fermionic
standard--like models. However, recently the search for flat 
$F$ and $D$ directions in the free fermionic models has been
systematized \cite{cceel2}. (See also \cite{dfset,gc}.) 
Thus revealing that the FNY
model does admit flat $F$ and $D$ solutions, but that these solutions
necessarily break the $U(1)_{Z^\prime}$ (as long as $U(1)_Y$ is preserved)
which is embedded in $SO(10)$.

In this paper we therefore utilize the 
important new developments of the tools needed to find
flat $F$ and $D$ solutions in the string models.
To achieve our goal of deriving the MSSM directly
from string theory, we have to incorporate the
set of fields needed to give mass to the fractionally
charged states into the possible solutions.
Similarly we will impose that the generated
VEVs also give large mass, of the order of
the string scale, to the additional color triplets
and electroweak doublets beyond the MSSM spectrum.
We should comment that in other semi--realistic
free fermionic models, the exotic states may also
receive large mass, but often only through high order 
nonrenormalizable terms,
and therefore their respective intermediate 
mass scales will be highly 
suppressed relative to the string scale. Therefore,
in the present letter, for the first time, we are able
to construct a string solution with solely the MSSM
spectrum in the $SU(3)\times SU(2)\times U(1)_Y$--charged
observable sector below the scale generated by the Fayet--Iliopoulos
term, a scale {\it on par} with the string scale.

\section{The model}

\subsection{Model construction}

The FNY string model \cite{fny} was constructed in the free fermionic
formulation \cite{fff}. It is generated by a set of
eight basis vectors of boundary conditions for all
the world--sheet free fermions. The first five vectors
consist of the NAHE set $\{{\bone}, \bs ,\bb_1,\bb_2,\bb_3\}$ \cite{nahe}.
In addition the basis contains three additional
vectors which are displayed in Table (\ref{fnymodel})
\beqn
 &\begin{tabular}{c|c|ccc|c|ccc|c}
 ~ & $\psi^\mu$ & $\chi^{12}$ & $\chi^{34}$ & $\chi^{56}$ &
        $\bar{\psi}^{1,...,5} $ &
        $\bar{\eta}^1 $&
        $\bar{\eta}^2 $&
        $\bar{\eta}^3 $&
        $\bar{\phi}^{1,...,8} $ \\
\hline
\hline
  ${\bb_4}$     &  1 & 1&0&0 & 1~1~1~1~1 & 1 & 0 & 0 & 0~0~0~0~0~0~0~0 \\
  ${\mbeta}$   &  1 & 0&0&1 & 1~1~1~0~0 & 1 & 0 & 1 & 1~1~1~1~0~0~0~0 \\
  ${\mgamma}$  &  1 & 0&1&0 &
                ${1\over2}$~${1\over2}$~${1\over2}$~${1\over2}$~${1\over2}$
              & ${1\over2}$ & ${1\over2}$ & ${1\over2}$ &
                ${1\over2}$~0~1~1~${1\over2}$~${1\over2}$~${1\over2}$~1 \\
\end{tabular}
   \nonumber\\
   ~  &  ~ \nonumber\\
   ~  &  ~ \nonumber\\
     &\begin{tabular}{c|c|c|c}
 ~&   $y^3{y}^6$
      $y^4{\bar y}^4$
      $y^5{\bar y}^5$
      ${\bar y}^3{\bar y}^6$
  &   $y^1{\omega}^6$
      $y^2{\bar y}^2$
      $\omega^5{\bar\omega}^5$
      ${\bar y}^1{\bar\omega}^6$
  &   $\omega^1{\omega}^3$
      $\omega^2{\bar\omega}^2$
      $\omega^4{\bar\omega}^4$
      ${\bar\omega}^1{\bar\omega}^3$ \\
\hline
\hline
$\bb_4$ & 1 ~~~ 0 ~~~ 0 ~~~ 1  & 0 ~~~ 0 ~~~ 1 ~~~ 0  & 0 ~~~ 0 ~~~ 1 ~~~ 0 \\
$\mbeta$  & 0 ~~~ 0 ~~~ 0 ~~~ 1  & 0 ~~~ 1 ~~~ 0 ~~~ 1  & 1 ~~~ 0 ~~~ 1 ~~~ 0 \\
$\mgamma$ & 0 ~~~ 0 ~~~ 1 ~~~ 1  & 1 ~~~ 0 ~~~ 0 ~~~ 1  & 0 ~~~ 1 ~~~ 0 ~~~ 0 \\
\end{tabular}
\label{fnymodel}
\eeqn
The choice of generalized GSO coefficients is:
\beqn
c\left(\matrix{\bb_4\cr
                                    \bb_j,\bbeta\cr}\right)&&=
-c\left(\matrix{\bb_4\cr
                                    {\bone}\cr}\right)=
-c\left(\matrix{\bbeta\cr
                                    {\bone}\cr}\right)=
c\left(\matrix{\bbeta\cr
                                    \bb_j\cr}\right)=\nonumber\\
-c\left(\matrix{\bbeta\cr
                                    \bgamma\cr}\right)&&=
c\left(\matrix{\bgamma\cr
                                    \bb_2\cr}\right)=
-c\left(\matrix{\bgamma\cr
                                    \bb_1,\bb_3,\bb_4,\bgamma\cr}\right)=
-1
\label{gsophases}
\eeqn
$(j=1,2,3),$
with the others specified by modular invariance and space--time
supersymmetry.
A property of the FNY model that distinguishes
it from the NAHE--based models of refs.\  \cite{eu,top} is the 
choice of pairings of the left-- and right--moving
real fermions from the set
$\{y,\omega\vert{\bar y},{\bar\omega}\}^{1,\cdots,6}$.
With the choice of pairing in FNY, the three left--moving
pairs $y^3y^6$, $y^1\omega^6$ and $\omega^1\omega^3$ are
complexified, and the remaining left--moving real fermions
from the set $\{y,\omega\}^{1,\cdots,6}$ are paired
with right--moving real fermions to produce Ising model operators.
In the right--moving sector 19 world--sheet fermions
are complex and generate the observable and hidden
four dimensional gauge groups.
We remark that both the choice of pairings as well
as that of the GSO phases affect nontrivially
the spectrum of the string models,
and consequently their specific phenomenological
characteristics \cite{slm}. 
This is an important remark,
keeping in mind that the eventual goal of string theory
is to learn how a particular vacuum is selected dynamically.

\subsection{Gauge group}

The gauge group before imposing the flat directions
consist of the universal observable $SO(10)$ sub--group,
$SU(3)_C\times SU(2)_L\times U(1)_C\times U(1)_L$,
generated by the five complex world--sheet fermions
${\bar\psi}^{1,\cdots,5}$; six observable horizontal,
flavor--dependent,
$U(1)$ symmetries $U(1)_{1,\cdots,6}$, generated by
$\{{\bar\eta}^1,{\bar\eta}^2,{\bar\eta}^3,{\bar y}^3{\bar y}^6,
{\bar y}^1{\bar\omega}^6,{\bar\omega}^1{\bar\omega}^3\}$,
respectively; and the unbroken sub--group 
$SO(4)\times SU(3)\times U(1)^4$ of the hidden $E_8$,
generated by ${\bar\phi}^{1,\cdots,8}$.
The weak hypercharge is given by 
\beq
U(1)_Y={1\over3}U(1)_C\pm{1\over2}U(1)_L
\label{weakhyper}
\eeq
where switching signs in eq.\  (\ref{weakhyper})
corresponds to flipping of the representations,
\beqn
     + &\leftrightarrow& -\\
 e_L^c &\leftrightarrow& N_L^c\nonumber\\
 u_L^c &\leftrightarrow& d_L^c\nonumber
\label{repflip}
\eeqn
This flip is equivalent to the flip between the straight
and flipped $SU(5)$ representations \cite{flipeq}. 
In the case of $SU(5)$
only the later choice is allowed as there are no adjoint representations
to break the non--Abelian gauge symmetry in the former.
In the case of
the standard--like models, as the GUT non--Abelian symmetry
is broken directly at the string level, this flip
is consistent with the requirement that the gauge
group can be broken to the Standard Model gauge group
in the effective low energy field theory.
We also note that under this $Z_2$ flip the Higgs representations
are also flipped $h\leftrightarrow{\bar h}$.

In the following we will show that the choice of the sign
in eq.\  (\ref{weakhyper}) also has interesting consequences
in terms of the decoupling of the exotic fractionally charged states.
The other combination of $U(1)_C$ and $U(1)_L$, which is
orthogonal to $U(1)_Y$ is given by
\beq
U(1)_{Z^\prime}=U(1)_C\mp U(1)_L
\label{u1prime}
\eeq
We will show that the choices of flat directions
in the model of ref.\  \cite{fny} forces either
$U(1)_Y$ or $U(1)_{Z^\prime}$ to be broken.
Therefore, in the phenomenologically viable case
we are forced to have only $SU(3)_C\times SU(2)_L\times U(1)_Y$
as the unbroken $SO(10)$ subgroup below the string scale.
While we have no dynamical mechanism that chooses
between these two distinct vacua, we find it to be
an interesting example of how string dynamics may
force the $SO(10)$ subgroup below the string scale
to coincide with the Standard Model gauge group.

\subsection{Matter spectrum}

The full massless spectrum of the model, together
with the quantum numbers under the right--moving
gauge group, are given in ref.\  \cite{fny}.
Here we give a brief summary. 
The sectors $\bb_1$, $\bb_2$ and $\bb_3$ correspond to the three
twisted sectors of the $Z_2\times Z_2$ orbifold model
and produce three generations in the 16 representation
of $SO(10)$ decomposed under
$SU(3)_C\times SU(2)_L\times U(1)_C\times U(1)_L$,
with charges under the horizontal symmetries.

The Neveu--Schwarz (NS) sector corresponds to the untwisted sector and 
produces in addition to the gravity and gauge multiplets three pairs of 
electroweak 
scalar doublets $\{h_1, h_2, h_3, {\bar h}_1, {\bar h}_2, {\bar h}_3\}$,
seven pairs of $SO(10)$ singlets with observable $U(1)$ charges, 
$\{\phi_{12},{\bar\phi}_{12},
   \phi_{23},{\bar\phi}_{23},
   \phi_{13},{\bar\phi}_{13},
   \phi_{56},{\bar\phi}_{56},
   \phi_{56}^\prime,{\bar\phi}_{56}^\prime,
   \phi_{4},{\bar\phi}_{4},
   \phi_{4}^\prime,{\bar\phi}_{4}^\prime\}$,
and three scalars that are singlets
of the entire four dimensional gauge group, $\phi_1,\phi_2,\phi_3$.

In the model of ref.\ \cite{fny}
the states from the NS sector and the sectors $\bb_1$, $\bb_2$ and $\bb_3$
are the only ones that transform solely under the observable,
$SU(3)_C\times S(2)_L\times U(1)_C\times U(1)_L\times U(1)_{1,\cdots,6}$
gauge group. The choice of GSO phase $c(\bb_4,1)=+1$ projects
all the massless states from the sector $\bb_4$ as well as
the space--time vector bosons from the sector $\mI=\bone +\bb_1+\bb_2+\bb_3$.

The sectors $\bb_j+2\mgamma$ and $\bb_j+2\mgamma+\mI$ with $j=1,\cdots,4$
produce $SO(10)$ singlet matter states in the 16 vector representation
of the hidden $SO(16)$ gauge group, decomposed under the final  
hidden group and are listed in Table I.
The sectors with some combination of
$\{\bone,\bb_1,\bb_2,\bb_3,\bb_4,\mbeta\}$ plus $\mgamma$
produce states that are $SU(3)_C\times SU(2)_L$
singlets, but are charged under $U(1)_Y$ or $U(1)_{Z^\prime}$.
Similar states arise also from the sectors
${\bone+\bb_1+\mbeta}+2\mgamma$ and $\mI+\mbeta+2\mgamma$.
These states arise due to the breaking of the $SO(10)$ symmetry,
by the basis vectors $\mbeta$ and $\mgamma$, and
carry fractional electric charge $\pm1/2$ or
fractional $U(1)_{Z\prime}$ charge. These are
exotic stringy states that do not fall into representations
of the original $SO(10)$ symmetry. One of the important
goals of our paper is to find the flat $F$ and $D$ solutions
that give {\it heavy mass} to all the exotic fractionally charged states.

The trilinear superpotential of the string model is
given by $W={\sqrt 2}g[W_1+W_2]$ where,
\beqn
W_1&=&\{({u_{L_1}^c}Q_1 +{N_{L_1}^c}L_1){\bar h}_1+
        ({d_{L_2}^c}Q_2 +{e_{L_2}^c}L_2) h_2+
        ({d_{L_3}^c}Q_3 +{e_{L_3}^c}L_3) h_3\nonumber\\
   &~&+h_2{\bar h}_1{\bar\phi}_{12}
      +{\bar h}_2{h_1}{\phi_{12}}+h_3{\bar h}_1{\bar\phi}_{13}
      +{\bar h}_3{h_1}{\phi_{13}}
      + h_3{\bar h}_2 {\phi_{23}}
      +{\bar h}_3{h_2}{{\bar \phi}_{23}}\nonumber\\
   &~&+\phi_{12}{\bar\phi}_{13}{\bar\phi}_{23}
      +{\bar\phi}_{12}\phi_{13}\phi_{23}
      +(\phi_4{\bar\phi}_4'+{\bar\phi}_4\phi_4')\phi_1\}\label{wone}\\
W_2&=&{1\over{\sqrt2}}\{H_1H_2\phi_4
      +(H_3H_4+H_5H_6){\bar\phi}_4
      +(H_7H_8+H_9H_{10})\phi_4'
      +H_{11}H_{13}{\bar\phi}_4'\nonumber\\
%   &~&\phantom{{1\over{\sqrt2}}}
   &~&\phantom{{0}}
      +(V_{41}V_{42} + V_{43}V_{44}){\bar\phi}_4
      +V_{45}V_{46}\phi_4+(V_{47}V_{48}+V_{49}V_{50}){\bar\phi}_4'
      +V_{51}V_{52}\phi_4'\}\nonumber\\
   &~&\phantom{{0}}
      +\{ H_{15}H_{16}\phi_{56}'
      + (H_{17}H_{18} + H_{21}H_{22}){\bar\phi}_{56}
      +H_{19}H_{20}{\bar\phi}_{56}' \} \nonumber\\
   &~&\phantom{{0}}
     +(V_{11}V_{12}+V_{13}V_{14}+V_{15} V_{17}+ V_{19}V_{20})\phi_{13}
       \nonumber\\
   &~&\phantom{{0}}
      +(V_{21}V_{22}+V_{23}V_{24}+V_{25} V_{27}+V_{29}V_{30})\phi_{12}
       \nonumber\\
   &~&\phantom{{0}}
      +V_{31}V_{32}{\bar\phi}_{23}+V_{33}V_{34}\phi_{23}
      +H_{29}H_{30}{\bar\phi}_{13}+H_{36}H_{37}\phi_{12}
%       \nonumber\\
\label{wtwo}\\
   &~&\phantom{{0}}
      +\{ (H_{31} H_{34} + V_{43} V_{45}) h_2 
      + V_{44} V_{46} {\bar h}_2
      + V_{47} V_{52} h_3 
      +(H_{38} H_{41} + V_{48} V_{51}) {\bar h}_{3} \}
.\nonumber
\eeqn

\subsubsection{Abelian anomaly}

{\hbox to 0.6truecm{\hfill}}With the choice of GSO projection coefficients in
eq.\  (\ref{gsophases})
the following $U(1)$ symmetries (from the entire set of 12
$\{U_C, U_L, U_{i=1,\, \, {\rm to}\, \, 6}, U_{j= H,\, 7,\, 8,\, 9}\}$)
are anomalous:
Tr${\, U_1=-24}$, Tr${\, U_2=-30}$, Tr${\, U_3=18}$,
Tr${\, U_5=6}$, Tr${\, U_6=6}$ and  Tr${\, U_8=12}$.
The total anomaly from all six of these Abelian symmetries
can be rotated into a single 
$U(1)_{\rm A}$, uniquely defined by 
\beq
         U(1)_{\rm A} \equiv c_A\sum_i \{\Tr Q^{(A)}_{i}\}U(1)_i,
\label{rotau1}
\eeq
with $c_A$ a normalization coefficient. In this case the single
anomalous Abelian symmetry becomes,
\beq
U_A\equiv -4U_1-5U_2+3U_3+U_5+U_6+2U_8.
\label{anomau1infny}
\eeq
After this rotation,
the universal Green--Schwarz relation invoked by modular invariance 
constraints removes all Abelian triangle anomalies except those involving
either one or three $U_A$ gauge bosons. 
That is, the five orthogonal, anomaly--free, Abelian combinations involving 
$U_{1,2,3,5,6,8}$, together with the other six anomaly free $U(1)$'s are
free of all gauge and mixed trilinear anomalies.

The standard anomaly cancellation
mechanism \cite{u1a} breaks $U_A$ and in the process generates an FI 
$D$--term,\footnote{Based on the arguments of \cite{jmr}, $M$--theory
does not appear to alter the form of the FI term. Instead an $M$--theory 
FI term should remain identical to the FI term obtained for a weakly--coupled
$E_8\times E_8$ heterotic string, independent of the size of 
$M$--theory's 11th dimension.} 
\beq
      \eps\equiv \frac{g^2_s M_P^2}{192\pi^2}\Tr Q_A\, ,
\label{fidt}
\eeq
where $g_{s}$ is the string coupling
and $M_P$ is the reduced Planck mass, 
$M_P\equiv M_{Planck}/\sqrt{8 \pi}\approx 2.4\times 10^{18}$ GeV. 
Spacetime supersymmetry is 
broken near the string scale by the FI $D_A$--term unless a set of scalar
VEVs, $\{\vev{\varphi_m}\}$, 
carrying anomalous charges $Q^{(A)}_m$ can contribute a compensating
$\vev{D_{A}(\varphi_m)} \equiv \sum_\alpha Q^{(A)}_m |\vev{\varphi_{m}}|^2$ 
term to cancel the FI term, i.e.,
\beq
\vev{D_{A}}= \sum_m Q^{(A)}_m |\vev{\varphi_{m}}|^2 
%+ \frac{g^2_s M_P^2}{192\pi^2}{\rm Tr}Q_{\rm A} = 0\,\, ,
+ \eps = 0\,\, ,
\label{daf}
\eeq
thereby restoring supersymmetry. 
A set of scalar VEVs satisfying eq.\ (\ref{daf})
is also constrained to keep $D$--flatness for all non-anomalous Abelian
$U_i$ symmetries as well,\footnote{Here we consider flat directions 
involving only non--Abelian singlet fields. In cases where non--trivial 
non--Abelian representations are also allowed to take on VEVs, generalized
non--Abelian $D$--flat constraints must also be imposed.}
\beq
\vev{D_i}= \sum_m Q^{(i)}_m |\vev{\varphi_{m}}|^2 = 0\,\, .
\label{dana}
\eeq

Each superfield $\Phi_{m}$ (containing a scalar field $\varphi_{m}$
and superpartner) 
in the superpotential imposes further constraints on the scalar VEVs. 
$F$--flatness will be broken (thereby destroying spacetime supersymmetry) at 
the scale of the VEVs unless,
\beq
\vev{F_{m}} = \vev{\frac{\partial W}{\partial \Phi_{m}}} = 0; \,\, \vev{W}  =0.
\label{ff}
\eeq

\subsubsection{Fractionally charged states}

{\hbox to 0.6truecm{\hfill}}By examining the fractionally charged 
states and the trilinear superpotential, it is
seen that {\it all} the fractionally charged states receive a heavy mass 
by giving a VEV to the neutral singlets 
${\bar\phi}_4,{\bar\phi}_4',{\phi}_4,\phi_4'$ \cite{fc}.
The additional trilinear terms,
$(\phi_4{\bar\phi}_4'+{\bar\phi}_4\phi_4')\phi_1$ 
impose an $F$--flatness constraint from eq.\ (\ref{ff}),
\beq
(\vev{\phi_4}\vev{{\bar\phi}_4'}+\vev{{\bar\phi}_4}\vev{\phi_4'})=0.
\label{fflatness}
\eeq

A $D$--flat solution to eqs.\ (\ref{daf}) and (\ref{dana})  
which also satisfies the $F$--flatness
constraints eq.\  (\ref{ff}) (and (\ref{fflatness}) in particular)
is given, for example, by the following set of fields
\beq
\{
\phi_{12}, \phi_{23}, {\bar\phi}_{56}, \phi_4, \phi_4^\prime,
{\bar\phi}_4,{\bar\phi}_4^\prime, H_{15}, H_{30}, 
H_{31}, H_{38} \}
\label{fdsol}
\eeq
with the VEVs
\beqn
&&{1\over 3}\vert\vev{\phi_{12}}\vert^2=
            \vert\vev{\phi_{23}}\vert^2=
            \vert\vev{{\bar\phi}_{56}}\vert^2=
  {1\over 2}\vert\vev{H_{15}}\vert^2=
  {1\over 3}\vert\vev{H_{30}}\vert^2=
            \vert\vev{H_{31}}\vert^2=
  {1\over 2}\vert\vev{H_{38}}\vert^2
\nonumber\\
&&\phantom{{1\over 3}\vert\vev{\phi_{23}}\vert^2} 
\equiv \vert \vev{\alpha} \vert^2\,\, ;
\nonumber\\
&& (\vert\vev{\phi_4}\vert^2+\vert\vev{\phi_4^\prime}\vert^2)-
(\vert\vev{{\bar\phi}_4}\vert^2+\vert\vev{{\bar\phi}_4^\prime}\vert^2)=
\vert \vev{\alpha} \vert^2\,\, ,
\label{fdsolval}
\eeqn
where $\vev{\alpha}$ is an overall scale 
(assumed to be real)\footnote{Together 
eqs.\ (\ref{fflatness}) and (\ref{fdsolval}) 
still allow degrees of freedom within the VEVs of the 
$\{\phi_{4},\phi^{'}_{4},\bar{\phi}_{4},\bar{\phi}^{'}_{4} \}$ fields.
VEV scales of orders lower than $\vert\vev{\alpha}\vert $ for these fields
would imply some mass scales of exotics being below the FI scale 
$\vert\vev{\alpha}\vert $.
However, naturalness of values would suggest that the VEVs of all of these
fields should be of the same scale as $\vert\vev{\alpha}\vert $.}
for the set of VEVs.\footnote{This flat direction is formed from 
a combination of VEVs, denoted $M_6$, $M_7$ and 
$R_{10}$ in \cite{cceel2}. 
$M_6$, $M_7$ and $R_{10}$ are individually $D$--flat
for each non--anomalous $U(1)$ and $R_{10}$ is responsible for
cancelling the FI term.}
 
Cancellation of the FI term by this $D$--flat solution 
specifies the scale $\vev{\alpha}$. Inserting the
flat direction VEVs and anomalous charge trace in (\ref{daf}) yields,  
\beq
\vev{D_{\rm A}}= -224 \vert \vev{\alpha} \vert^2
+ \frac{g^2_{s} \MP^2 \,\,  1344}{192\pi^2} = 0\,\, .
\label{adeta}
\eeq
At $M_U = 2.5\times 10^{16}$ GeV, the unified couplings have a value
$\alpha_U \equiv g^2_{physical}/(4\pi )\approx 1/24$. This
corresponds to $g_{s}\approx .5$, since
$g_{physical}= \sqrt{2} g_{s}\approx .72$.
Thus, the FI scale for the FNY model is 
\beq
\vert \vev{\alpha} \vert \approx 7\times 10^{16}\,\,\, {\rm GeV}.
\label{adetb}
\eeq

In ref.\  \cite{cfn2},
the space of $F$-- and $D$--flat directions
that decouple the fractionally charged states will be studied in more detail.
The following important remarks are, however, in order.
Following the holomorphic gauge--invariant polynomial method developed in 
refs.\  \cite{dfset} and \cite{cceel2} 
or the matrix method of ref.\ \cite{gc}, 
it is possible to classify the complete space
of allowed $D$--flat directions. 
When the non--Abelian singlet fields allowed to take on VEVs 
are restricted to those that do not break the weak--hypercharge,
it is observed that
all the possible
$D$--flat solutions (that cancel the FI term) 
contain fields that break the $U(1)_{Z^\prime}$ symmetry, 
eq.\  (\ref{u1prime}). 
In other words, no directions that are
$D$--flat for all non--anomalous Abelian symmetries 
and are formed solely from
non--Abelian singlet fields uncharged under both  
$U(1)_{Y}$ and $U(1)_{Z'}$,
make a contribution to the anomalous $D$--term
that can cacell the FI--term. Specifically,
anomalous $D$--term cotribution is $0$ for all
of these direction.  

We then have the very interesting situation in which
the $U(1)_{Z^\prime}$ is necessarily broken 
in the $D$--flat string vacuum. 
The FNY model therefore
presents the first example where the string consistency
constraints force the $SO(10)$ symmetry to be broken
to $SU(3)_C\times SU(2)_L\times U(1)_Y$ by singlet VEVs, 
rather than 
$SU(3)\times SU(2)_L\times U(1)_C\times U(1)_L$. Possible 
generalization of this constraint for VEVs of non--Abelian states
will be examined in \cite{cfn2}.

We should remark, however, that we also anticipate
that there exist, phenomenologically unviable,
$D$--flat singlet solutions in which the weak--hypercharge
is broken and the $U(1)_{Z^\prime}$ remains
unbroken. The existence of such solutions
will also be investigated in ref.\  \cite{cfn2}.
At present we do not know what is the mechanism,
or the reason, that selects the $U(1)_Y$--preserving
vacuum over the $U(1)_Z^\prime$--preserving vacuum.

The $D$--flat direction (\ref{fdsolval}) was found
to be $F$--flat to {\it all} orders in the string--based 
superpotential \cite{cfn2}.
$F$--flatness is broken in a 
field--theoretic gauge--invariant superpotential by generic terms of the form
\beqn
(\phi_{1,2,3})^{m= 0,\, 1}(\phi_{4} \bar{\phi}_{4})^{n}\,\, &;&\quad
(\phi_{1,2,3})^{m= 0,\, 1}(\phi^{'}_{4} \bar{\phi}^{'}_{4})^{n}\,\, ;  
\nolabel\\        
(\phi_{1,2,3})^{m= 0,\, 1}(\phi_{4} \bar{\phi}^{'}_4)^{n}\,\, &;&\quad
(\phi_{1,2,3})^{m= 0,\, 1}(\phi^{'}_4 \bar{\phi}_4)^{n}\,\, ,
\label{fbtb}
\eeqn
where $n$ is any positive integer.
However, string world--sheet selection rules impose 
strong constraints on allowed superpotential terms beyond
gauge invariance. Here, for example, all (\ref{fbtb}) terms
are forbidden by the requirement that no more than $N-4$ (for $N\ge 4$)
NS fields can appear in an $N^{\rm th}$ order term \cite{wsc,nahew5}.
All fields in (\ref{fbtb}) are NS class.      

The next observation we wish to make 
is to recall that we have a $Z_2$
ambiguity in the definition of the weak--hypercharge, eq.
(\ref{weakhyper}). Choosing the positive sign
imposes the decoupling condition on the fractionally charged
states, eq.\  (\ref{fflatness}). Choosing the
negative sign in the definition of the weak--hypercharge,
eq.\  (\ref{weakhyper}), flips between
the exotic fractionally charged states and the 
electrically neutral exotic states with fractional
$U(1)_{Z^\prime}$ charge.
There is a class of exotic states from the sectors
$\bone+\bb_1+\mbeta+2\mgamma$ and $\bI+\mbeta+2\mgamma$ 
which is invariant under this flip. 
These states do not have $U(1)_C$
charge and are either $SU(2)_L$ singlets with $U(1)_L=\pm1$,
or $SU(2)_L$ doublets with $U(1)_L=0$. In both cases
they carry fractional electric charge $\pm1/2$.
With the positive sign in the weak--hypercharge 
definition all the states with fractional electric charge
gain mass at the cubic level of the superpotential
by the VEVs of $\phi_4$, $\phi_4^\prime $,$\bar\phi_4$,
$\bar\phi_4^\prime$. With the negative
sign definition of the weak--hypercharge
the states $\{H_{15},H_{16},H_{17},H_{18},H_{19},H_{20},
H_{21},H_{22},H_{23},H_{24},H_{25},H_{26},H_{27},H_{28}\}$
in Table 2 of ref. \cite{fny},
carry fractional electric charge $\pm1/2$ and are not coupled
to  $\phi_4$, $\phi_4^\prime $,$\bar\phi_4$,
$\bar\phi_4^\prime$ at the cubic level of the superpotential.
The states from the sectors
$\bone+\bb_1+\mbeta+2\mgamma$ and $\mI+\mbeta+2\mgamma$
are, of course, still electrically charged and
can receive heavy mass from the VEVs of 
$\phi_4$, $\phi_4^\prime $,$\bar\phi_4$,
$\bar\phi_4^\prime$. We can contemplate
other $F$--flat and $D$--flat solutions
that will make all the fractionally charged states
super--heavy. However, we see that with the
negative sign definition of the weak--hypercharge
this will be far more difficult to achieve.
The discussion above illustrates how 
also the $Z_2$ ambiguity in the weak--hypercharge
definition is broken by the choices of $F$-- and $D$--flat
solutions. 

\subsubsection{Exotic triplet/anti--triplet pair}

{\hbox to 0.6truecm{\hfill}}The massless spectrum of the FNY model contains
in addition to the exotic fractionally charged
states, one pair of $SU(3)_C$ triplets in vector--like
representation ($H_{33},H_{40}$) with fractional
$U(1)_{Z^\prime}$ charge and several pairs of electroweak
Higgs doublets. 
To show that with the $F$-- and $D$--flat solution, 
eq.\  (\ref{fdsolval}), the spectrum below the string scale 
indeed coincides with that of the MSSM, 
we have to show that in this vacuum these additional states
receive mass near the string scale,  
with only one pair of Higgs doublets remaining light.

A mass term,
\beq
m_{trip}= \frac{A_5\vev{\phi_{23}}\vev{H_{38}}\vev{H_{31}}}{M^2_{S}}
= \vev{\alpha}\frac{\sqrt{2} A_5\vev{\alpha}^2}{M^2_{S}},
\eeq
for the additional color triplet pair appears from the quintic superpotential 
term,
\beq
H_{33}H_{40}H_{31}H_{38}\phi_{23}
\label{qmassterm}
\eeq
While this triplet mass is generated at the 
FI scale $= \vev{\alpha} \approx 7\times 10^{16}$ GeV,
we estimate the additional factor in $m_{trip}$, which includes 
the five--point string amplitude $A_5$
(defined here to not include the $1/M_S^2$ factor)
for this specific fifth--order nonrenormalizable coupling,
will contribute a suppression factor on the order of 
$\sim ({1}/{10}-1/100)$. 
The five--point string amplitude includes a 
world--sheet integral $I_2$.
Similar world--sheet integrals have been computed 
for other fifth order superpotential terms, both in different NAHE-- 
\cite{nahew5} and non--NAHE--based \cite{cceeli,cew} free fermionic 
models. The other worldsheet integral values were all 
found to be of the same order and our estimate of $m_{trip}$ 
here assumes our $I_2$ value is comparable to those others.
(Note that the triplet mass ``suppression factor''  
$(\sqrt{2} A_5 \vev{\alpha}/M_{S})^2$ would appear to be of order one.)
While the exotic triplet mass may lie slightly below the 
string/unification scale of $M_U=2.5\times 10^{16}$ GeV,
it appears sufficiently close to $M_U$ so as not to 
significantly affect the running of the MSSM couplings.   
We emphasize that the numerical estimate of the masses
arising from the singlet VEVs should be regarded only as
illustrative. The important result is the generation of
mass terms for all the states beyond the MSSM,
near the string scale. The actual masses of the extra
fields may be spread around the $M_U$ scale, thus
inducing small threshold corrections that are still
expected to be compatible with the low energy
experimental data.

\subsubsection{Exotic doublets}

%hb states: (hb1, hb2, hb3, H34)
%h  states: ( h1,  h2,  h3, H41)

{\hbox to 0.6truecm{\hfill}}We next turn to the analysis of the Higgs mass 
spectrum generated by the $F$-- and $D$--flat solution eq.\ (\ref{fdsolval}).
The FNY model contains four $h_{up}$-class MSSM $SU(2)$ doublets
\beq
\{ \hb_1,  \hb_2, \hb_3, \hb_4\equiv H_{34} \}
\label{hup1}
\eeq
and four $h_{down}$--class doublets,
\beq
\{ h_1, h_2, h_3, h_4\equiv H_{41} \}\, .  
\label{hdn1}
\eeq
A mass matrix $M$ yielding doublet mass terms,
\beq
 ( h_1, h_2, h_3, h_4) M ( \hb_1,  \hb_2, \hb_3, \hb_4)^{T}\, , 
\eeq
results from the flat direction VEVs.   
This mass matrix   
\begin{equation}
\label{massmatrix}
M_{h_i,\hb_j}=\left ( 
\begin{array}{cccc}
0 & g \vev{\phi_{12}} & 0  &  0 \\
0 & 0                 & 0  & g \vev{H_{31}} \\ 
0 & g \vev{\phi_{23}} & 0  &  0 \\
0 & 0                 & g\vev{H_{38}} & 
\frac{A_5 \vev{\phi_{23}}\vev{H_{38}}\vev{H_{31}}}{M^2_{s}}
\end{array} \right ).
\label{mar1}
\end{equation}
with $\vev{\phi_{12}}= \sqrt{3}\vev{\alpha}$, 
     $\vev{\phi_{23}}= \vev{\alpha}$, 
     $\vev{H_{31}}  = \vev{\alpha}$, and
     $\vev{H_{38}}= \sqrt{2}\vev{\alpha}$.
The zeros in the matrix hold to all orders in the stringy superpotential.
This can be proven simply by gauge--invariance constraints and 
the $N-4$ NS rule. 
The NS rule eliminates three non--zero mass terms, that would  
appear in a field--theoretic model:
\beqn 
M_{h_1 \bar{h}_4} &=& \vev{\phi_{12} H_{31}} A_4/M_s\,\, ;\quad 
M_{h_3 \bar{h}_4}= \vev{\phi_{23} H_{31}} A^{'}_4/M_s\,\, ;
\label{massft1}\\
M_{h_4 \bar{h}_2} &=& \vev{\phi_{23} H_{38}} A^{''}_4/M_s \,\, .   
\label{massft2}
\eeqn
The field--theoretic $M_{h_1 \bar{h}_4}$ and $M_{h_3 \bar{h}_4}$
values appear in the same ratio as
$M_{h_1 \bar{h}_2}$ and $M_{h_3 \bar{h}_2}$ (provided $A_4 = A^{'}_4$).
Hence, we note that the field--theoretic and string models involve 
the same eigenstates and eigenmasses of $h_1$ and $h_3$.  

To determine the $h$ and $\hb$ mass eigenstates and eigenvalues 
we evaluate the eigenstates and eigenvalues of $M M^{\dag}$ and
$M^{\dag} M$ respectively. We find there is exactly one massless 
Higgs--like eigenvalue pair:
\beq
h'_{3}  = \frac{1}{2}(- h_1 + \sqrt{3} h_3) \qandq
\hb_1\qwithq m^2_{h'_3}= m^2_{\hb_1}= 0\,\, ; \label{h0}
\eeq
and three pairs with string scale masses:
\beqn
 h'_1&=&   \frac{1}{2}(\sqrt{3} h_1 + h_3) 
\qandq \hb_2 \qwithq
m^2_{h'_1}= m^2_{\hb_2} =  4 g^2 \vev{\alpha}^2\,\, ; \label{hm1}\\
&&\nonumber\\
h'_{2}&=&c_{2'}[- (A^2 + B^2 + \sqrt{A^4 + 6 A^2 B^2 + B^4}\, ) \, h_2 
                                                     + (2 A B) \, h_4] 
\qand\label{h24l}\\ 
\hb'_4&=&\bar{c}_{4'}[(2 \sqrt{2} A B)\, \hb_3 - 
                (A^2 - B^2 + \sqrt{A^4 + 6 A^2 B^2 + B^4}\,) \, \hb_4]
\qwith\nonumber\\
m^2_{h'_2}&=& m^2_{\hb'_4}=(3 A^2 + B^2 - 
    \sqrt{-8 A^4 + (3 A^2 + B^2)^2})/2\,\, ;
\nonumber\\
& &\nonumber\\
h'_{4}&=& c_{2'}[(2 A B)\, h_2 
               + (A^2 + B^2 + \sqrt{A^4 + 6 A^2 B^2 + B^4}\, )\, h_4]
\qand\label{h24h}\\
\hb'_3 &=& \bar{c}_{4'}[
(A^2 - B^2 + \sqrt{A^4 + 6 A^2 B^2 + B^4}\, )\, \hb_3 
+ (2 \sqrt{2} A B)\, \hb_4 ]
\qwith\nonumber\\
m^2_{h'_4}&=& m^2_{\hb'_3}= 
(3 A^2 + B^2 + \sqrt{-8 A^4 + (3 A^2 + B^2)^2})/2 \,\, ;
\nonumber
\eeqn
where $A\equiv g \vev{\alpha}$, 
$B\equiv \frac{\sqrt{2} A_5 \vev{\alpha}^3}{M^2_{s}}$,
and $c_{2'}$ and $\bar{c}_4$ are normalization constants.

Thus, $h'_2$, $\hb'_3$,  $h'_4$, and $\hb'_3$ obtain masses of 
${\cal O}(\vev{\alpha})$, along with the fractionally charged states. 
Note that the Higgs spectrum
is simplified in the $B<< A$ limit,
(i.e., no mass mixing term between $H_{41}$ and $H_{35}$), the 
$h_{2,4}$ and $\hb_{3,4}$ mass eigenstates and eigenvalues reduce to
\beqn
h_2 & {\rm and} &  \hb_4\quad {\rm with} \quad
m^2_{h_2}= m^2_{\hb_4}= A^2 \equiv g^2 |\vev{\alpha}|^2
\label{m24}\\
h_4 & {\rm and} &  \hb_3\quad {\rm with} \quad
m^2_{h_4}= m^2_{\hb_3}= 2 A^2 \equiv 2 g^2 |\vev{\alpha}|^2\,\, .
\label{m42}
\eeqn
Hence, consistent with MSSM physics, we see that the flat direction of
eq.\ (\ref{fdsolval}) produces in the low energy effective field theory
a single pair of massless Higgs doublets 
above the supersymmetry breaking scale.

\subsubsection{Exotic singlets \& hidden sector states}

The FNY model contains 
16 non--Abelian singlet states with fractional electric charge, 
44 hypercharge--neutral singlets, and three uncharged states. 
All 16 fractionally charged
states receive induced masses from renormalizable superpotential terms
involving VEVs of (\ref{fdsol}). 
$14+1$ of the remaining $44+3$ singlets 
similarly receive masses from third order
terms, while an additional four singlets receive mass at fifth order.
$26+2$ singlets remain massless through at least sixth order. 

Approximately two--thirds of the 34 non--Abelian hidden sector states
become massive from 
third (all 4 fractionally charged states plus 8 other states), 
fourth (6 states), or fifth order terms (4 states). 
One $SU(3)_H$ triplet/anti--triplet pair, six $SU(2)_H$ doublets,
and four $SU(2)^{'}_H$ doublets remain massless through at least sixth order. 

The appearance of exotic singlets and hidden non-Abelian states  
uncharged under the SM, still has significant phenomenological implications.
In particular, when such states appear in a string model, 
we should expect them to interact with SM states via the extra $U(1)$ charges.
Surprisingly, for the particular flat direction considered we find that,
after decoupling of the massive states from the low energy theory,
no such interactions appearing in the effective Yukawa  
produced by the third order non-renomalizable superpotential and at least  
fourth through sixth order superpotential terms\footnote{Higher order contributions
will be investigated in \cite{cfn2}.} containing one to three VEVs, respectively.
Interestingly, all hidden sector terms in the effective Yukawa appear with
$\vev{\alpha}/M_s$ factors. That is, all hidden sector ``non-suppressed'' third order terms
are removed when massive states are decoupled.

\vskip .2truecm

{
\def\H#1{H_{#1}}
\def\V#1{V_{#1}}
\def\h#1{h_{#1}}
\def\hp#1{h^{'}_{#1}}
\def\hb#1{\bar{h}_{#1}}
\def\L#1{L_{#1}}
\def\Q#1{Q_{#1}}
\def\dc#1{d^{c}_{#1}}
\def\uc#1{u^{c}_{#1}}
\def\ec#1{e^{c}_{#1}}
\def\Nc#1{N^{c}_{#1}}

\def\p#1{{\phi}_{#1}}
\def\pp#1{\phi^{'}_{#1}}
\def\pb#1{\bar{\phi}_{#1}}
\def\ppb#1{\bar{\phi}^{'}_{#1}}
%       g_s \sqrt{2} hv01 [ S019   hv09 +  UR03  b05 ]  
%       g_s \sqrt{2} hv06 [ S016   hv07 +  UR01  b02 ]  
%       \lambda_5  <S014> <S018>  b01   hv07  UR02  
%       g_s \sqrt{2}   T003 [ S001   S004 +  S002  S003 ]  
%       g_s \sqrt{2}   S006   S031   S060  
%       g_s \sqrt{2}   S009   S039   S040  
%       \lambda_4  <S055> S031   H042   H062     
%       \lambda_4  <S054> S039   H005   H066    
%       \lambda_5  <S002> <S055> S031   H043  H060    
%       \lambda_5  <S002> <S054> S039   H006  H064   
%        \lambda_4     \frac{<\H{31}>}{M_s} \H{19} \V{19} \V{37} +    
%        \lambda^{'}_4 \frac{<\H{30}>}{M_s} \V{31} \H{25} \V{35} +   
%        \lambda_5     \frac{<\ppb{4}> <\H{31}>}{M_{s}^{2}}\H{19} \V{15} \V{40} +   
%        \lambda^{'}_5 \frac{<\ppb{4}> <\H{30}>}{M_{s}^{2}}\V{31}  \H{23} \V{39}  

\beqn
W_{(MSSM)}&:&
        g_s \sqrt{2}\,\, \hb{1} [ \Q{1} \uc{1} + \L{1} \Nc{1} ]  +
        g_s \sqrt{2}\,\, \cos{\sqrt{3}}\,\, \hp{3} [ \Q{3} \dc{3} + \L{3} \ec{3} ] 
\label{wyssm}\\
W_{(Singlet)}&:&
        g_s \sqrt{2}\,\,  \p{1} [ \p{4} \ppb{4} + \pp{4} \pb{4} ] + 
        g_s \sqrt{2}\,\,  \ppb{56} \H{19} \H{20} +
        g_s \sqrt{2}\,\,  \pb{23} \V{31} \V{32}  
\label{wysig}\\
W_{(Hidden)}&:&
\lambda_4\,\,     \frac{<\alpha>}{M_s} \H{19} \V{19} \V{37} +    
\lambda^{'}_4\,\, \frac{\sqrt{3}<\alpha>}{M_s} \V{31} \H{25} \V{35} + 
\label{wyhid}\\
& &  
\lambda_5\,\,  \frac{<\ppb{4}> <\alpha>}{M_{s}^{2}}\H{19} \V{15} \V{40} +   
\lambda^{'}_5\,\,
    \frac{\sqrt{3}<\ppb{4}><\alpha>}{M_{s}^{2}}\V{31}  \H{23} \V{39}  
\nonumber
\eeqn}
{\noindent}Effective Yukawa terms generated by third through sixth order 
superpotential terms containing one to three fields with VEVs, respectively. 
 
\section{Comments}

Exotic MSSM states, many carrying fractional electric charge 
are a generic feature of many classes of string models. Most
of these, if they remain massless down to the electroweak scale,
signify unphysical phenomenology, thereby disallowing a model containing them. 
``String--selection rules'' can make decoupling of 
dangerous exotic fields from the low energy effective field theory 
difficult. They often forbid several superpotential terms, 
otherwise allowed by gauge invariance, that could 
generate large mass for an exotic via couplings with 
flat direction VEVs \cite{cceelw}.

Intermediate scale MSSM exotics are more phenomenologically viable
than electroweak scale exotics. However, intermediate scale exotics
will generally alter the running of the MSSM couplings and shift
the unification scale away from the MSSM projected value of
$M_U\approx 2.5\times 10^{16}$ GeV. Further,
one may argue that intermediate mass scales
for MSSM exotics require additional ad hoc fine tuning, and 
therefore are not very attractive. 

In this letter we have presented a string model wherein
it is actually possible to decouple all SM--charged MSSM exotics from the 
effective field theory, giving mass to these fields at the 
Fayet--Iliopoulos (i.e., anomalous $U(1)$) scale, 
which is very near the string scale.
Thus, we have found a string model consistent with both 
projected unification of the MSSM couplings at
$M_{U}\approx 2.5\times 10^{16}$ GeV
and the conjecture that the string scale $M_S$ may in fact coincide with
$M_{U}$. 
This is the first string model that we are aware of with these properties.  

The particular flat direction of the FNY model chosen herein is, in fact,
not the model's only flat direction that decouples all MSSM exotics.
A more complete set of flat directions that likewise perform this task will be
presented in \cite{cfn2}. 
Detailed analysis of the physics of these different flat directions, 
including MSSM mass hierarchies, will be performed. 
Two difficulties with the example 
flat direction presented herein are that the flat direction does not yield
an effective $\mu$ term at any order in the superpotential;  
and it gives (through at least sixth order terms) a top-$\nu_{\tau}$ 
universality without a seesaw mechanism, 
resulting in an ${\cal{O}}(100$ GeV$)$ neutrino.
Our search for phenomenologically superior flat directions, 
more consistent with the MSSM, will be discussed in \cite{cfn2}. 

\section{Acknowledgments}
This work is supported in part
by DOE Grants No. DE--FG--0294ER40823 (AF)
and DE--FG--0395ER40917 (GC,DVN).

%=========================================================================
%\half, \mhalf, \thrd, \mthrd,  \frth, \mfrth, \tfrth, \mtfrth
% include $...$ in definition for these

\def\half{${\textstyle{1\over 2}}$}
\def\mhalf{${-\textstyle{1\over 2}}$}
\def\thrd{${\textstyle{1\over3}}$}
\def\mthrd{${-\textstyle{1\over3}}$}
\def\frth{${\textstyle{1\over4}}$}
\def\mfrth{${-\textstyle{1\over4}}$}
\def\tfrth{${\textstyle{3\over4}}$}
\def\mtfrth{${-\textstyle{3\over4}}$}

\begin{center} 
%                1 345678901 23 4567
\begin{tabular}{|c|rrrrrrrrr|cc|rrrr|}
\hline
%  1     2             3              4    5      6    7     8      9    10    11    12       13     14    15    16    17
State&$(C,L)$&$Q_C$&$Q_L$&$Q_1$&$Q_2$&$Q_3$&$Q_4$&$Q_5$&$Q_6$&$SO(4)$&$SU(3)$&$Q_H$&$Q_7$&$Q_8$&$Q_9$   \\
\hline
\hline
$V_{41}$ &(1,1)&  0 &  1 &  0 &  \half &  0 &  \half &  0 & \mhalf &(1,1)&(1)&  0 &  \half &  0 & \mhalf \\
$V_{42}$ &(1,1)&  0 & -1 &  0 & \mhalf &  0 &  \half &  0 &  \half &(1,1)&(1)&  0 & \mhalf &  0 &  \half \\
$V_{43}$ &(1,1)&  0 &  1 &  0 & \mhalf &  0 &  \half &  0 & \mhalf &(1,1)&(1)&  0 & \mhalf &  0 &  \half \\
$V_{44}$ &(1,1)&  0 & -1 &  0 &  \half &  0 &  \half &  0 &  \half &(1,1)&(1)&  0 &  \half &  0 & \mhalf \\
$V_{45}$ &(1,2)&  0 &  0 &  0 & \mhalf &  0 & \mhalf &  0 &  \half &(1,1)&(1)&  0 &  \half &  0 & \mhalf \\
$V_{46}$ &(1,2)&  0 &  0 &  0 &  \half &  0 & \mhalf &  0 & \mhalf &(1,1)&(1)&  0 & \mhalf &  0 &  \half \\
$V_{47}$ &(1,1)&  0 &  1 &  0 &  0 &  \half &  \half &  \half &  0 &(1,1)&(1)&  0 & \mhalf &  0 &  \half \\
$V_{48}$ &(1,1)&  0 & -1 &  0 &  0 & \mhalf &  \half & \mhalf &  0 &(1,1)&(1)&  0 &  \half &  0 & \mhalf \\
$V_{49}$ &(1,1)&  0 &  1 &  0 &  0 & \mhalf &  \half &  \half &  0 &(1,1)&(1)&  0 &  \half &  0 & \mhalf \\
$V_{50}$ &(1,1)&  0 & -1 &  0 &  0 &  \half &  \half & \mhalf &  0 &(1,1)&(1)&  0 & \mhalf &  0 &  \half \\
$V_{51}$ &(1,2)&  0 &  0 &  0 &  0 & \mhalf & \mhalf &  \half &  0 &(1,1)&(1)&  0 & \mhalf &  0 &  \half \\
$V_{52}$ &(1,2)&  0 &  0 &  0 &  0 &  \half & \mhalf & \mhalf &  0 &(1,1)&(1)&  0 &  \half &  0 & \mhalf \\
\hline
\hline
\end{tabular}
\end{center}
\bigskip
\noindent Table I.a Fractionally charged states of class A with electric charges $\pm$\half.
($C$ and $L$ in column three denote the observable sector $SU(3)_C$ and $SU(2)_L$.)

\begin{center}
%                1 345678901 23 4567
\begin{tabular}{|c|rrrrrrrrr|rr|rrrr|}
\hline
\hline
%  1     2             3              4    5      6    7     8      9    10    11    12       13     14    15    16    17
State&$(C,L)$&$Q_C$&$Q_L$&$Q_1$&$Q_2$&$Q_3$&$Q_4$&$Q_5$&$Q_6$&$SO(4)$&$SU(3)$&$Q_H$&$Q_7$&$Q_8$&$Q_9$   \\
\hline
$H_1$    &(1,1)&  \tfrth &  \half &  \frth & \frth &  \frth & \mhalf &  \half &  0 &(2,1)&(1)& \tfrth &  \frth &  0 &  0 \\
$H_2$    &(1,1)& \mtfrth & \mhalf & \mfrth & \mfrth & \mfrth & \mhalf & \mhalf &  0 &(2,1)&(1)& \mtfrth & \mfrth &  0 & 0\\
$H_3$    &(1,1)&  \tfrth &  \half &  \frth &  \frth &  \frth &  \half &  \half &  0 &(1,1)&(1)& \mtfrth & \mfrth & \mhalf &  \half \\
$H_4$    &(1,1)& \mtfrth & \mhalf & \mfrth & \mfrth & \mfrth &  \half & \mhalf &  0 &(1,1)&(1)&  \tfrth &  \frth &  \half & \mhalf \\
$H_5$    &(1,1)&  \tfrth &  \half &  \frth & \frth &  \frth &  \half & \mhalf &  0 &(1,1)&(1)& \mtfrth & \mfrth &  \half & \mhalf \\
$H_6$    &(1,1)& \mtfrth & \mhalf & \mfrth & \mfrth & \mfrth &  \half &  \half &  0 &(1,1)&(1)&  \tfrth &  \frth & \mhalf &  \half \cr
$H_7$    &(1,1)&  \tfrth &  \half &  \frth& \mfrth & \mfrth & \mhalf &  0 &  \half &(1,1)&(1)&\mtfrth & \mfrth &  \half &  \half \\
$H_8$    &(1,1)& \mtfrth & \mhalf & \mfrth &  \frth &  \frth & \mhalf &  0 & \mhalf &(1,1)&(1)& \tfrth &  \frth & \mhalf & \mhalf \\
$H_9$    &(1,1)&  \tfrth &  \half &  \frth& \mfrth & \mfrth & \mhalf &  0 & \mhalf &(1,1)&(1)&\mtfrth & \mfrth & \mhalf & \mhalf \\
$H_{10}$ &(1,1)& \mtfrth & \mhalf & \mfrth &  \frth &  \frth & \mhalf &  0 &  \half &(1,1)&(1)& \tfrth &  \frth &  \half &  \half \cr
$H_{11}$ &(1,1)&  \tfrth &  \half &  \frth & \mfrth & \mfrth &  \half &  0 & \half  &(1,2)&(1)&  \tfrth &  \frth &  0 &  0 \\
$H_{13}$ &(1,1)& \mtfrth & \mhalf & \mfrth &  \frth &  \frth &  \half &  0 & \mhalf &(1,2)&(1)& \mtfrth & \mfrth &  0 &  0 \\
\hline
\hline
\end{tabular}
\end{center}
\bigskip
\noindent Table I.b Fractionally charged states of class B with electric charges $\pm$\half.

\vfill
\eject

\begin{center}
%                1 345678901 23 4567
\begin{tabular}{|c|crrrrrrrr|rr|rrrr|}
\hline
\hline
%  1     2             3              4    5      6    7     8      9    10    11    12       13     14    15    16    17
State&$(C,L)$&$Q_C$&$Q_L$&$Q_1$&$Q_2$&$Q_3$&$Q_4$&$Q_5$&$Q_6$&$SO(4)$&$SU(3)$&$Q_H$&$Q_7$&$Q_8$&$Q_9$   \\
\hline
$H_{33}$ &(3,1)& \mfrth & \mhalf & \mfrth & \mfrth & \mfrth &  0 &  0 & \mhalf &(1,1)&(1)&  \tfrth & \mfrth &  \half &  0 \\
$H_{40}$ &($\bar 3$,1)&  \frth &  \half &  \frth & \mfrth & \mfrth &  0 & \mhalf &  0 &(1,1)&(1)& \mtfrth &  \frth &  \half &  0 \\
\hline
\hline
\end{tabular}
\end{center}
\bigskip
\noindent Table I.c Fractionally charged states of class C with electric charges $\pm$\thrd.

\begin{center}%
                
\begin{tabular}{|l|l|}
\hline
\hline
%  1     2    
Sector& States\\
\hline
${\bone+\bb_1+\bbeta+2\bgamma}$    & $V_{41}$ to $V_{46}$ \\   
${\mI+\bbeta+2\bgamma}$            & $V_{47}$ to $V_{52}$ \\
$\pm$ $\bgamma$                    & $H_1$ to $H_2$       \\
${\mI\pm\bgamma}$                  & $H_3$ to $H_6$       \\
${\bone+\bb_4\pm\bgamma}$          & $H_7$ to $H_{10}$    \\
${\mI+\bone+\bb_4\pm\bgamma}$      & $H_{11}$ to $H_{13}$ \\
${\bb_3+\beta\pm\bgamma}$          & $H_{33}$             \\
${\bb_1+\bb_2+\bb_4+\bbeta}$       & $H_{40}$             \\
\hline
\hline
\end{tabular}
\end{center}
\bigskip
\noindent Table I.d Fractionally charged state sectors.

%======================== REFERENCES =====================================
%=========================================================================
%atbib

\vfill\eject

\bigskip
\medskip

\bibliographystyle{unsrt}

\begin{thebibliography}{99}

\bibitem{mth} {E. Witten, \NPB{471}{96}{135};\\
D.V. Nanopoulos, ``M-Phenomenology,'' CTP-TAMU-42/97, ACT-15/97,
[hep-th/9711080].}

\bibitem{fny} A.E. Faraggi, D.V. Nanopoulos, and K. Yuan, \NPB{335}{90}{347}.

\bibitem{fc} A.E. Faraggi, \PRD{46}{92}{3204}.

\bibitem{cceel2} {G. Cleaver, M. Cveti\v c, J.R. Espinosa, L. Everett, and
  P. Langacker, \NPB{525}{98}{3}, [hep-th/9711178];
``Flat Directions in Three Generation String Models,'' 
UPR-0784T, [hep-th/9805133].}

\bibitem{anoma} For general discussions of anomalous $U(1)$ in string models
see, e.g.,\\ 
T. Kobayashi and H. Nakano, \NPB{496}{97}{103}, [hep-th/9612066];\\
G.B. Cleaver, \NPBPS{62A-C}{98}{161} [hep-th/9708023];\\
G.B. Cleaver and A.E. Faraggi, UFIFT-HEP-97-28, UPR-0773-T, [hep-ph/9711339];\\
A.E. Faraggi, \PLB{426}{98}{315}, [hep-ph/9801409];\\
L.E.~Ib\'{a}\~{n}ez, R. Rabadan, and A.M. Uranga, [hep-th/9808139];\\  
P. Ramond, Proceedings of Orbis Scientiae '97 II, 
Dec. 1997, Miami Beach, [hep-ph/9808488];\\
W. Pokorski and G.G. Ross, [hep-ph/9809537]
and references contained within each.

\bibitem{u1a} M. Dine, N. Seiberg and E. Witten, \NPB{289}{86}{585};\\
                J. Atick, L. Dixon and A. Sen, \NPB{292}{87}{109}.

\bibitem{so10sig} I. Antoniadis, J. Ellis, J. Hagelin and D.V. Nanopoulos
                \PLB{231}{89}{65};\\ 
                I. Antoniadis. G.K. Leontaris and J. Rizos,
                                \PLB{245}{90}{161};\\
               J.L. Lopez, D.V. Nanopoulos and K. Yuan, \NPB{399}{93}{654};\\
               \AEF, \PLB{339}{94}{223}.

\bibitem{eu} \AEF, \PLB{278}{92}{131}.

\bibitem{dfset} {M. Luty, \PRD{53}{96}{3399}, [hep-th/9506098];\\
T. Gherghetta, C. Kolda, and S. Martin, \NPB{468}{96}{37}, [hep-th/9510370];\\
E. Dudas, C. Grojean, S. Pokorski, and C.A. Savoy, \NPB{481}{96}{85}, 
[hep-ph/9606383];\\
P. Binetruy, N. Irges, S. Lavignac, P. Ramond, \PLB{403}{97}{38}, 
[hep-ph/9612442];\\
N. Irges and S. Lavignac \PLB{424}{98}{293}, [hep-ph/9712239].}

\bibitem{gc} {G. Cleaver, ``Mass Hierarchy and Flat Directions in 
String Models,'' UPR-0795T. Proceedings of Orbis Scientiae '97 II,
Dec. 1997, Miami Beach.}

\bibitem{fff} {I. Antoniadis, C. Bachas, and C. Kounnas, \NPB{289}{87}{87};\\
               H. Kawai, D.C. Lewellen, and S.H.-H. Tye, \NPB{288}{87}{1}.}

\bibitem{nahe} A.E. Faraggi and D.V. Nanopoulos, \PRD{48}{93}{3288}.

\bibitem{top} A.E. Faraggi, \PLB{274}{92}{47}.

\bibitem{slm} A.E. Faraggi, \NPB{387}{92}{239}; [hep-th/9708112].

\bibitem{flipeq} S.M. Barr, \PLB{112}{82}{219};\\
                 J.P. Derendinger, J.E. Kim and D.V. Nanopoulos, 
                \PLB{139}{84}{170};\\
                I. Antoniadis, J. Ellis, J.S. Hagelin and D.V. Nanopoulos,
                \PLB{194}{87}{231}.

\bibitem{jmr} {J. March-Russell, ``The Fayet-Iliopoulos term in Type-I 
string theory and M-theory,'' IASSNS-HEP-97/128, CERN-TH/98-198,
[hep-ph/9806425].}

\bibitem{cfn2} G. Cleaver, A.E. Faraggi and D.V. Nanopoulos,
papers in preparation. 

\bibitem{nahew5} {S. Kalara, J. L\'opez, and D.V. Nanopoulos, 
\PLB{245}{90}{421};
\NPB{353}{91}{650};\\ 
A.E. Faraggi, \NPB{487}{97}{55}.}

\bibitem{wsc} {D. Bailin, D. Dunbar, and A. Love, \PLB{219}{89}{76};\\
J. Rizos and K. Tamvakis, \PLB{262}{91}{227}.}

\bibitem{cceeli} {G. Cleaver, M. Cveti\v c, J.R. Espinosa, L. Everett, and
  P. Langacker, \PRD{57}{98}{2701}, [hep-ph/9705391].}  

\bibitem{cew} {M. Cveti\v c, L. Everett, J. Wang, ``Units and Numerical 
Values of the Effective Couplings in Perturbative Heterotic String Theory,''
UPR-0810-T, [hep-ph/9808321].}

\bibitem{cceelw} {G. Cleaver, M. Cveti\v c, J.R. Espinosa, L. Everett, 
  P. Langacker and J. Wang, 
``Physics Implications of Flat Directions in Free Fermionic Superstring
Models I: Mass Spectrum and Couplings,''
CERN-TH/98-243, UPR-0811-T, IEM-FT-177/98, UM-TH-98/17, 
[hep-ph/9807479];  
``Physics Implications of Flat Directions in Free Fermionic Superstring
Models II: Renormalization Group Analysis,''
UPR-0814-T, UM-TH-98/18, [hep-ph/9811355].}

%****************************************************************************
\end{thebibliography}

\vfill\eject
\end{document}